# Phase diagrams of polarized ultra-cold gases on attractive-U Hubbard ladders

**E.A. Burovski[1], R.Sh. Ikhsanov[1], A.A. Kuznetsov[1] and M.Yu. Kagan[1,2]**

[1] National Research University Higher School of Economics, 101000 Moscow, Russia
[2] P.L. Kapitza Institute for Physical Problems, Russian Academy of Sciences, 119334, Moscow, Russia

rihsanov@hse.ru

**Abstract**. We consider a quasi-one-dimensional model of a two-component Fermi gas at zero temperature on one, two and three-leg attractive-U Hubbard ladders. We construct the grand canonical phase diagram of a two-component spin-polarized gas. We find that the structure of the phase diagram of the attractive-U Hubbard model for two and three leg ladders significantly differs q from the structure of the phase diagram of a single chain. We argue that the single chain model is a special case, and that multichain ladders display qualitative features of the 1D-to-3D crossover, observed in experiments with trapped ultracold gases.

## 1. Introduction

Recent advances in experimental techniques of creation and manipulation of strongly correlated systems using ultra-cold gases in traps with varying forms and effective dimensionality of the trapping potential [1, 2, 3, 4] stimulates an interest in studying quasi-one-dimensional Fermi gases at ultralow temperatures. In such systems a motion of the particles in two spatial dimensions out of three is frozen out by dimensional quantization. The possibility of experimental engineering of one-dimensional and quasi-one-dimensional geometries opens up a unique opportunity for studying polarized superfluids, or, more specifically, the one-dimensional analog [5] of the long-elusive Fulde-Ferrell-Larkin-Ovchinnikov polarized superfluid phase [6, 7], and the dimensional crossover between the strictly one-dimensional (1D) and three-dimensional (3D) phenomena.

An important feature of the experimental studies of ultracold gases is that the (strongly interacting) atomic clouds are subject to an external confining potential. This leads to a non-trivial spatial structure of the atomic cloud, which is directly observable in either time-of-flight experiments or, recently, with an *in situ* imaging. For polarized Fermi gases, the spatial structure of the atomic cloud was found to be significantly different for quasi 1D cigar-shaped magnetic traps [4] and 3D spherically symmetric traps [3]. A recent experimental effort, Ref. [1], investigates a crossover between the 1D and 3D in an array of cigar-shaped tubes with varying the intertube tunneling amplitudes.

In this Paper, we use a quasi-one-dimensional model of a two-component Fermi gas at zero temperature to map out the grand canonical phase diagram of the uniform system. With the local density approximation, the grand canonical phase diagram provides the shell structure of the density profile of a trapped gas, which can be compared to experimental measurements. Specifically, we consider an attractive Hubbard model on two- and three-leg ladders and compare the behavior of the model to the known case of a strictly 1D single chain lattice [8–11]. We concentrate on studying the shell structure of the density profile, and leave aside the details of the algebraic ordering in 1D versus the true long-range ordering in 3D. We use the numerically exact density matrix renormalization group (DMRG) simulations. This way, our work complements a recent study [12], which used a mean field approximation.

## 2. Model and methods

We consider the attractive-U Hubbard model ($U < 0$) defined by the Hamiltonian,

$$H_0 = -t\sum_{i,\sigma}\left(c^+_{i,\sigma}c_{i+1,\sigma} + h.c.\right) + U\sum_i n_{i\uparrow}n_{i\downarrow} , \quad (1)$$

where $c^+_{i,\sigma}$ is the creation operator for a fermion with spin $\sigma$ at site $i$, $n_{i,\sigma} = c^+_{i,\sigma}c_{i,\sigma}$ is the number of particles operator; summations run over $L\times w$ sites of a $w$-leg ladder lattice of length $L$. $U$ is the onsite Hubbard interaction between fermions with opposite spins, and $t$ is the hopping amplitude, which we set $t = 1$ without loss of the generality.

We only consider zero temperature ($T = 0$) and compute the ground state energies using the DMRG method [13, 14], as implemented in the ALPS package [15].

In the canonic ensemble, the ground state energy $E\left(N_\uparrow, N_\downarrow\right)$ of the model (1) is the function on the particle numbers for spin-up and spin-down fermions, $N_\uparrow$ and $N_\downarrow$, respectively. For convenience, we also define the total number of particles, $N_\Sigma = N_\uparrow + N_\downarrow$, and total polarization, $N_\Delta = N_\uparrow - N_\downarrow$. It is enough to consider only the case $N_\Sigma < L\times w$ because of the particle-hole symmetry of the model (1) and $N_\Delta > 0$. The case $N_\Delta < 0$ corresponds to a trivial relabelling of up and down spins: $\uparrow\leftrightarrow\downarrow$.

Given the energies, we can transform to the grand canonic ensemble, which corresponds to the change of variables: the convenient variables being total chemical potential $\mu$, and the effective magnetic field, $h$, are given by the derivatives of the energy with respect to $N_\Sigma$ and $N_\Delta$, respectively: $\mu = \left(\dfrac{\partial E}{\partial N_\Sigma}\right)_{N_\Delta}$, $h = \left(\dfrac{\partial E}{\partial N_\Delta}\right)_{N_\Sigma}$.

Since our main interest is the grand canonical phase diagram, we compute the ground state energies in the canonical ensemble, and approximate the derivatives with finite differences. The possible phases of the model (1) are well documented in the literature (see, e.g. [9] and references therein). The following phases are of interest:

(i) vacuum (V), which corresponds to $N_\uparrow = N_\downarrow = 0$;

(ii) equal densities (ED), $N_\uparrow = N_\downarrow$;

(iii) the partially polarized phase (PP) with $N_\uparrow > N_\downarrow$;

(iv) the fully polarized phase ($FP_1$), $N_\downarrow = 0$.

In the presence of the lattice, there are additional phases, which correspond to a full band fillings (e.g., $FP_2$: $N_\uparrow = L\times w$ and $N_\downarrow = 0$). The ED phase is expected to have the Bardeen-Cooper-Schrieffer (BCS) type of algebraic ordering, while the PP phase is expected to display the the FFLO type order, where the superconding correlations decay algebraically in real space and are modulated with the modulation length given by the mismatch of the Fermi momenta [5].

To locate the phase boundaries, we compare the energies in different phases:

a) The boundary between ED and PP. At the boundary we have $N_\uparrow = N_\downarrow$, therefore the chemical potential and the effective magnetic field are given by:

$$\mu = \left(\frac{\partial E}{\partial N_\Sigma}\right)_{N_\Delta} = \frac{E(N_\uparrow + 1, N_\downarrow + 1) - E(N_\uparrow, N_\downarrow)}{2}, \quad h = \left(\frac{\partial E}{\partial N_\Delta}\right)_{N_\Sigma} = \frac{E(N_\uparrow + 1, N_\downarrow - 1) - E(N_\uparrow, N_\downarrow)}{2}.$$

On the $\mu - h$ plane, the boundary is parameterized by $N_\uparrow$.

b) The boundary between ED and the vacuum is the related to the binding energy of a single pair, and therefore on the $\mu - h$ plane the boundary is nothing but the horizontal line $\mu = E(1,1)/2$.

c) The boundary between the vacuum and $FP_1$ phase is defined by the bottom of the single-particle band, so that the boundary is given by the straight line $\mu = -h + E(1,0)$.

d) The boundary between the partially polarized phase PP and fully polarized phase $FP_1$. The boundary is defined by:

$$\mu = \left( \frac{\partial E}{\partial N_\Sigma} \right)_{N_\Delta} = \frac{E(N_\uparrow + 1,1) - E(N_\uparrow, 0)}{2}, \; h = \left( \frac{\partial E}{\partial N_\Delta} \right)_{N_\Sigma} = \frac{E(N_\uparrow - 1,1) - E(N_\uparrow, 0)}{-2}.$$

On the $\mu - h$ plane, the boundary is parameterized by $N_\uparrow$.

e) The boundary between the fully polarized phases $FP_1$ and $FP_2$ is given by the straight line $\mu = -h - E(1,0)$.

## 3. Results and Discussion

We map the grand canonical phase diagram for the attractive-U Hubbard model on the chain and two- and three-leg ladders with up to 120 sites and varying the Hubbard interaction parameter up to $U/t = -7$. In our DMRG simulations we keep up to 400 states.

Our numerical results are summarized in Fig. 1(a)-(c). For $w = 1$ our results agree with the Bethe Ansatz based calculations [9, 16, 17]. For the two-leg ladders ($w = 2$) our results agree with the DMRG simulations of Ref [8].

The grand canonical phase diagrams, Fig. 1, define the shell structure of the atomic clouds, which are directly observable in experiments. In the local density approximation, a weak spin-independent confining potential $V(x)$ is equivalent to replacing $h \to h - V(x)$, which graphically corresponds to a vertical slice in Fig. 1 [9].

Several features of the phase diagrams are interesting and important. First, the non-trivial boundaries (ED-PP, line 1, and PP-FP1, line 4) are non-monotonic for $w > 1$ (for two and three-leg ladders). This is a purely lattice effect due to filling of the higher bands at finite values of the filling fraction [8]. Experimentally, the most relevant part of the phase diagram is the vicinity of the multicritical point *O*, since it corresponds to the low filling fraction, i.e. to the continuum limit. A qualitative feature is the slope of the line 1, which separates the equal-density phase and the partially polarized phase PP: for $d\mu/dh < 0$ in the vicinity of the multicritical point *O*, the PP phase occupies the center of the trap, and the ED phase is either absent, or occupies the wings of the atomic cloud [18]. This behavior has been observed for strictly one-dimensional problem ($w = 1$) in Bethe Ansatz calculations in the continuum space [18] and on the lattice [9], also numerically in the DMRG calculations [9] and experimentally in tight cigar-shaped traps [4].

For both $w = 2$ and $w = 3$, we find that the slope of the phase boundary is positive, $d\mu/dh > 0$, see Fig 1(b) and 1(c). In the local density approximation, this translates into a different scenario for the shell structure of the atomic cloud, where the structure of the cloud is inverted: the center of the trap has equal densities, while the wings can be partially polarized. This is closer to the symmetric 3D trap results of Ref. [3], which observed the phase separation between the ED core and FP wings.

While our calculations are only done for two- and three-leg ladders, it is thus natural to conjecture that this scenario is generic, and that the strictly 1D, single-chain is special in this respect.

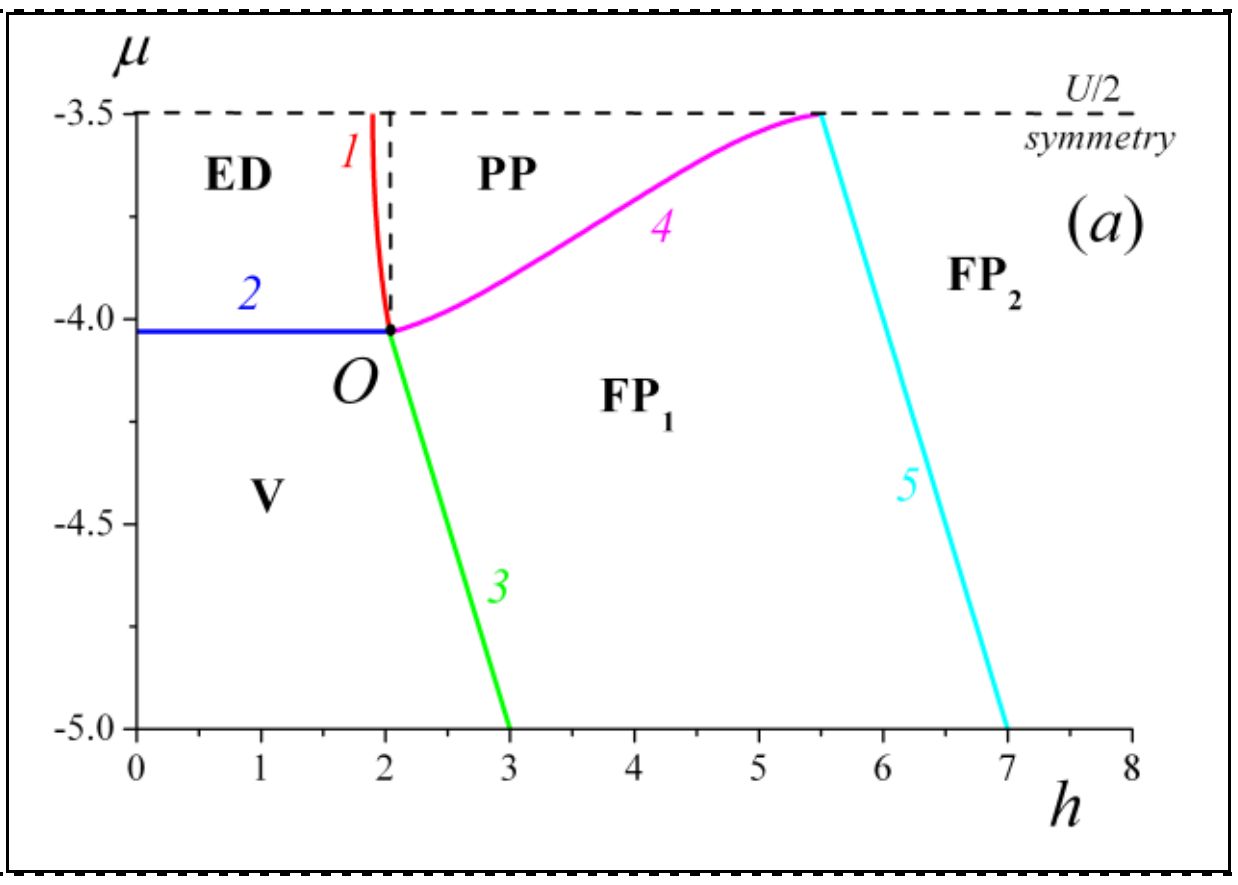

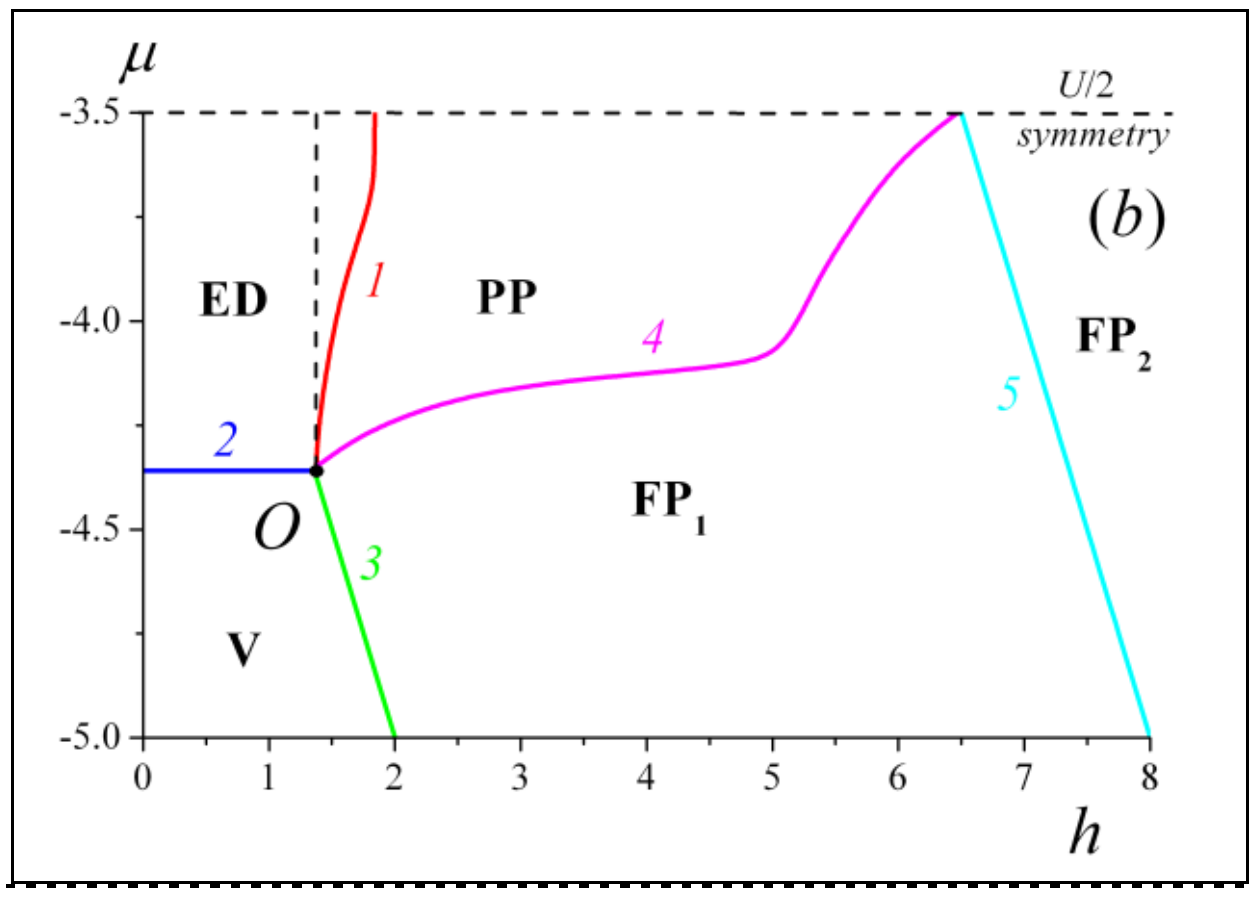

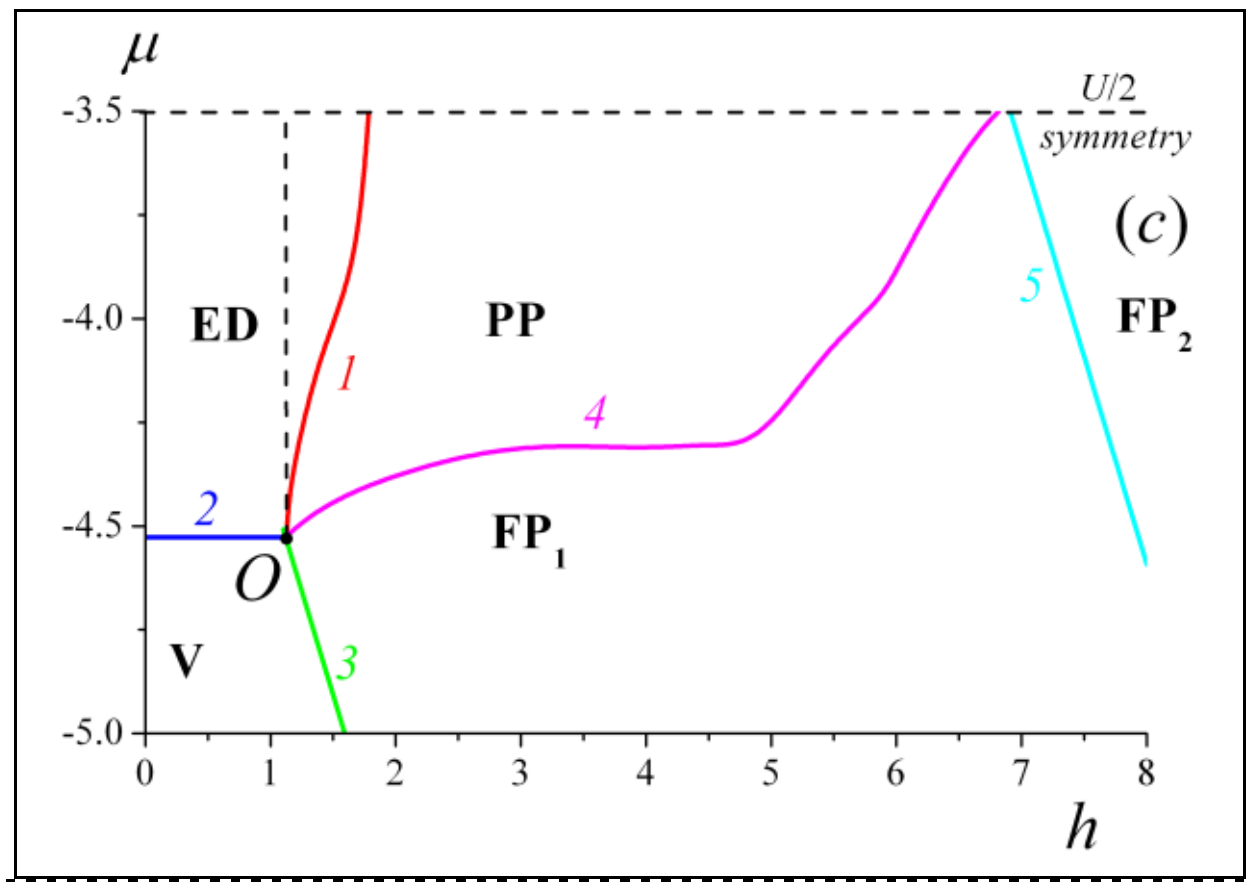

**Figure 1.** (color online) Grand-canonical phase diagrams of the attractive-U Hubbard model for $U = -7t$ and varying number of legs $w$. We only show the $\mu < U/2$, $h > 0$ part of the phase diagram since the $h < 0$ part is symmetric with respect to the $h \to -h$ symmetry, and $\mu > U/2$ is symmetric with respect to the particle-hole symmetry of the Hubbard model (1). Solid lines are simulation results, the vertical dashed line is guide to an eye only. See the text for the discussion. (a) $w = 1$, $L = 40$; (b) $w = 2$, $L = 40$; (c) $w = 3$, $L = 40$.

## 4. Conclusions and Outlook

We perform numerically exact DMRG simulations of the attractive-U Hubbard model on a single chain and on two- and three-leg ladders, which complement the mean-field studies of Ref. [12]. We construct the grand canonical phase diagrams, which allow "reading off" the shell structure of the atomic cloud in the experiments with quasi-one-dimensional polarized ultracold gases. We find that the structure of the phase diagram for multi-leg ladders has qualitative differences from the single chain case, which is consistent with recent experimental results for arrays of weakly coupled 1D tubes [1]. It would be interesting to study the effect of the varying interchain hopping amplitude to mimic the experimental setup, which we reserve for a future study.

## 5. Acknowledgments

E.A.B. and M.Yu.K. acknowledge the support of the Academic Fund Program at the National Research University Higher School of Economics (HSE) in 2018−2019 (grant № 18-05-0024) and by the Russian Academic Excellence Project "5-100". The work of R.S.I. was financially supported by joint Russian-Greek projects RFMEFI61717X0001 and T4ΔPΩ-00031 "Experimental and theoretical studies of physical properties of low-dimensional quantum nanoelectronic systems". We are grateful also to RFBR grant № 17-02-00135.